# High-gain vortex transfer via activated forbidden transitions in molecular magnets

Fan Meng, Hao Zhu, Xin-Yao Huang[†], Guo-Feng Zhang[*]

*School of Physics, Beihang University, Beijing 102206, China*

**Abstract**

Vortex light has garnered considerable interest in recent years owing to its distinctive properties and broad application potential. In this paper, we employ a conventional three-level ladder-type configuration in molecular magnets to realize high-gain vortex light transfer by enabling otherwise forbidden transitions. We demonstrate that the intensity and phase of the generated vortex signal field are governed by the detuning of the probe field and the strength of the control field, and that the topological charges of the incident and generated fields obey a well-defined algebraic relation during the vortex transfer process. Furthermore, we show that, in molecular magnets, the Autler-Townes splitting (ATS) effect yields a larger vortex signal field gain than electromagnetically induced transparency (EIT) over a broad parameter range in nonlinear three-wave mixing. This result suggests that the widely accepted view of EIT as the dominant enhancer of nonlinear optical effects may not hold universally. In addition, we revisit previous studies and re-examine the characterization of vortex beam transfer efficiency, emphasizing the need for a more careful interpretation in multi-beam interaction systems. Leveraging the long spin coherence and microwave transitions of molecular magnets, our results may enable quantum information transfer, storage, computing, and radar imaging in solid-state platforms.

## 1. Introduction

In recent years, molecular magnetic materials have attracted significant research interest [1]. These materials consist of nanoscale molecules featuring a magnetic metal center coordinated by nonmagnetic atoms (ligands) [2]. Depending on the ligand composition, their molecular sizes can range from 1 to 100 nanometers. Owing to the presence of ligands, intermolecular interactions within the crystal lattice are significantly weaker than intramolecular interactions. As a result, these systems are often treated as non-interacting single-molecule magnets (SMMs). Despite possessing large spins compared to free electrons, these magnetic molecules can exhibit microscopic quantum phenomena—such as quantum coherence, tunneling, and discrete magnetization steps—in the behavior of macroscopic systems [2-4]. Thus, they bridge classical macroscopic behavior and quantum nanoscale phenomena [5].

Advances in the synthesis of molecular magnetic materials have enabled precise control over molecular dimensions, paving the way for in-depth investigations into their magnetic properties. A variety of remarkable phenomena have been experimentally observed in molecular magnet crystals, including abrupt magnetization reversals

[†] Corresponding author. E-mail: xinyaohuang@buaa.edu.cn

[*] Corresponding author. E-mail: gf1978zhang@buaa.edu.cn

(avalanches) [3], avalanche-associated electromagnetic radiation [6,7], and magnetic deflagration—wherein the avalanche propagates through the crystal at a constant velocity [8,9]. Molecular magnets have been proposed as promising sources of coherent microwave radiation [10]. Their strong resonant absorption of electromagnetic radiation has been confirmed via electron spin resonance (ESR) experiments [11,12], forming the basis for investigating the interaction between molecular magnets and electromagnetic (including optical) and acoustic waves. A range of important optical phenomena have been explored in molecular magnets, including electromagnetically induced transparency (EIT) [13], four-wave mixing (FWM) [14], giant Kerr nonlinearities and microwave solitons [15], as well as the excitation and propagation of acoustic waves [16,17]. Owing to their long magnetic relaxation times at cryogenic temperatures and nanoscale dimensions, molecular magnets hold significant promise for applications in magnetic storage, information processing, and quantum computing [18-20].

Vortex light has also garnered widespread interest as a novel form of structured light [21]. It carries a spiral phase structure described by the factor $e^{il\varphi}$, resulting in a helical phase distribution. Here, $l$ denotes the topological charge, which may be an integer or non-integer, with its sign indicating the direction of the helical wavefront; $\varphi$ represents the azimuthal angle in the transverse plane of the beam. Such beams carry orbital angular momentum (OAM) of $l\hbar$ per photon and exhibit a characteristic doughnut-shaped intensity profile with a central dark core—corresponding to a phase singularity where the phase is undefined. Although scalar vortex light carrying OAM is perhaps the most widely studied example, control over other degrees of freedom is also attracting increasing attention, offering a route to higher-dimensional forms of structured light [22]. Examples include spatial vector vortex light with spatially dependent polarization patterns [23,24], spatiotemporal vortex light carrying transverse OAM in the spatiotemporal domain [25,26], and spatiotemporal vector vortex light with nonseparable spatiotemporal-polarization states [27]. Vortex light has demonstrated significant utility in optical communication [28,29] and quantum information science [30,31], and has also proven valuable in optical tweezers [32], optical testing [33], laser material processing [34], and microscopy imaging [35]. Moreover, numerous intriguing and important phenomena have been observed in the interaction between vortex light and matter [36-52]. In particular, frequency conversion of vortex beams through light-matter interactions has been widely reported [45-52], though most studies focus on the visible spectral range. Extending such studies to lower-frequency regimes, such as the microwave domain, is also of considerable significance. This is due to the widespread use of such frequencies in radar imaging [53], target detection [54], macroscopic manipulation [55], and wireless data transmission [56].

Molecular magnets provide an excellent platform for investigating microwave optical vortices. In molecular magnet crystals, the frequency of optical vortices can be tuned by varying the intensity of an external direct current (DC) magnetic field; furthermore, nonlinear processes in these materials allow the generation of microwave optical vortices across multiple frequencies. Most previous studies of nonlinear effects in molecular magnets have been conducted under the EIT regime [13-15,17]. EIT is a quantum interference phenomenon that renders an otherwise opaque medium transparent [57]. It has been shown that EIT can significantly enhance nonlinear optical processes while suppressing linear absorption [58]. In contrast, the Autler-Townes splitting (ATS) effect, which is spectrally similar to EIT [59,60], has received comparatively little attention. Although efficient nonlinear frequency mixing of Gaussian beams via the ATS effect has been demonstrated in atomic ensembles [61], the corresponding nonlinear frequency conversion of vortex beams in molecular magnets remains largely unexplored. Given the flexible tunability of the host matrix and the extended ground-state spin coherence times [62], molecular magnets may offer practical advantages over atomic systems.

In this work, we demonstrate high-gain vortex light transmission via a three-level ladder-type system in molecular magnets. We show that both the intensity and phase of the generated vortex signal field are governed by the detuning of the probe field and the amplitude of the control field, and that the topological charges of the vortex beams obey a specific algebraic relation in the nonlinear three-wave mixing (TWM) process. It is worth noting that, in natural three-level ladder atomic systems, second-order nonlinear sum-frequency generation is forbidden because the potential energy possesses strict spatial inversion symmetry. Molecular magnets, however, provide a route to overcome this restriction through explicit symmetry breaking in the effective spin Hamiltonian. This symmetry breaking renders the eigenstates no longer pure spin-projection states. The resulting state mixing relaxes the conventional magnetic-dipole selection rules, thereby enabling TWM signal beam generation through otherwise forbidden transitions in the ladder configuration. The activation of such forbidden transitions has been experimentally demonstrated in the nuclear-spin-free chromium complex $[Cr(C_3S_5)_3]^{3-}$ [63]. Inspired by this result, we construct a closed-loop configuration for nonlinear three-wave mixing using three mixed spin eigenstates of this molecular magnet. Furthermore, we compare the roles of EIT and ATS in vortex beam transfer and demonstrate that ATS offers a significantly enhanced signal gain over a broad parameter range. Our findings broaden the understanding of vortex light-matter interactions, offering a pathway to overcome forbidden transitions and realize efficient vortex beam conversion. This approach may also facilitate practical implementations of quantum information storage, optical communication, quantum computing, and related applications in solid-state systems.

## 2. Theoretical model

As illustrated in Fig. 1, we consider a three-level ladder-type system in which a weak probe field $\Omega_p$ couples the transition $|1\rangle \leftrightarrow |2\rangle$, a strong control field $\Omega_c$ drives the transition $|2\rangle \leftrightarrow |3\rangle$, and the generated signal field $\Omega_s$ couples the transition $|1\rangle \leftrightarrow |3\rangle$. In the axially symmetric case, the spin Hamiltonian is invariant under rotations about the symmetry axis and therefore commutes with the corresponding spin-projection operator, i.e., $\left[\hat{H}_0, \hat{S}_z\right] = 0$. As a consequence, its eigenstates can be labeled by the well-defined magnetic quantum number $M_s$. Transitions driven by a linearly polarized microwave magnetic field transverse to the symmetry axis obey the magnetic-dipole selection rule $\Delta M_s = \pm 1$. A transition such as $|M_s = -3/2\rangle \leftrightarrow |M_s = +3/2\rangle$ is therefore forbidden in the pure-spin limit because $\Delta M_s = 3$. In molecular magnet, the rhombic magnetic anisotropy and the transverse static magnetic field explicitly break the axial symmetry of the effective spin Hamiltonian. Consequently, $\left[\hat{H}_0, \hat{S}_z\right] \neq 0$, and $M_s$ is no longer a good quantum number. The resulting mixed-spin eigenstates contain components connected by the magnetic-dipole operators $\hat{S}_x$ and $\hat{S}_y$, allowing the nominally forbidden transition $|1\rangle \leftrightarrow |3\rangle$ to acquire a finite magnetic-dipole matrix element. We refer to this mechanism as symmetry-breaking activation of the forbidden transition. The violation of these rules can induce novel quantum phenomena, offering a new avenue for exploring second-order nonlinear effects in quantum systems. In our model, we employ the nuclear-spin-free complex $[Cr(C_3S_5)_3]^{3-}$, synthesized by M.S. Fataftah *et al.*, which exhibits low magnetic anisotropy and a high rhombicity (E/D). These characteristics result in strongly mixed $M_s$ levels at low magnetic fields, thereby enabling appreciable intensity in nominally forbidden transitions [63]. Importantly, the absence of nuclear spins suppresses nuclear spin-mediated decoherence and avoids hyperfine coupling that would otherwise complicate spectral interpretation, ensuring that the observed transitions arise purely from electron spin ($S = 3/2$) dynamics. In this system, we assign the upper state $|3\rangle$ to $|M_s = +1/2\rangle$, the intermediate state $|2\rangle$ to $|M_s = -1/2\rangle$, and the lower state $|1\rangle$ to $|M_s = -3/2\rangle$ (see Fig. S5 in Ref. [63]). The $M_s$ labels used above indicate only their adiabatic high-field parentage and should not be interpreted as pure spin states.

We assume that magnetic molecules are non-interacting, and that the system can be described by a single-molecule spin Hamiltonian. Each molecule is subjected to a DC magnetic field oriented perpendicular to its easy axis, while an alternating current (AC) magnetic field is applied along the easy axis. We define the $z$-axis as the easy anisotropy axis and the $x$-axis as the direction of the DC magnetic field. Thus, the total Hamiltonian of the system [13] can be written as $\hat{H} = \hat{H}_0 + \hat{V}$ ($\hbar = 1$), where

$$\hat{H}_0 = D\hat{S}_z^2 + E\left(\hat{S}_x^2 - \hat{S}_y^2\right) + g_x \mu_B H_0 \hat{S}_x, \tag{1}$$

$$\hat{V} = \frac{\mu_B}{2} \sum_{j=p,c,s} \hat{\mathbf{S}} \cdot \left(\mathbf{g} \cdot \mathbf{H}_j\right) e^{-i\omega_j t + i\mathbf{k}_j \cdot \mathbf{r}} + \text{H.c.}. \tag{2}$$

Here, $\hat{\mathbf{S}}$ denotes the spin operator, with $\hat{S}_x, \hat{S}_y$, and $\hat{S}_z$ representing its projections along the $x$, $y$, and $z$ axes, respectively. $E\left(\hat{S}_y^2 - \hat{S}_x^2\right)$ denotes the transverse anisotropic energy term. $H_0$ represents the DC magnetic field, while $D$, $\mathbf{g}$, and $\mu_B$ correspond to the longitudinal anisotropy constant, Landé factor, and Bohr magneton, respectively. $\hat{V}$ is the interaction operator between the magnetic molecule and the AC magnetic field (incident field). $\mathbf{H}_j$ represents the amplitude of the AC magnetic field. Each molecule exhibits a doublet energy structure, which arises either from the DC magnetic field alone or from the combined influence of the DC field and transverse anisotropy. For the selected complex, the experimentally determined spin-Hamiltonian parameters are $D = 0.260(7)\text{cm}^{-1}$, $E = -0.080(2)\text{cm}^{-1}$, $g_x = 1.935(1)$, $g_y = 2.020(1)$, $g_z = 2.060(1)$ and $H_0 = 2000G$ [63]. Under these conditions, different $M_S$ components can be strongly mixed at low magnetic fields, such that the actual eigenstates are no longer pure basis states but rather linear combinations of multiple spin states. Consequently, the conventional selection rule $\Delta M_S = \pm 1$ is relaxed, and otherwise forbidden transitions acquire nonzero dipole transition matrix elements through state mixing, enabling coherent driving by pulsed microwave fields.

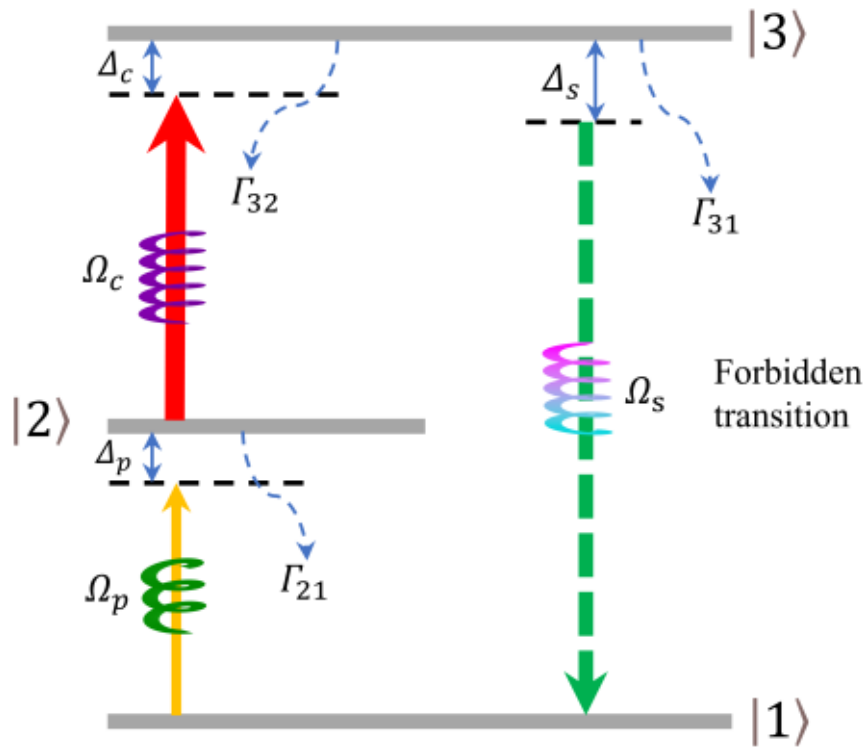

**Fig. 1.** A three-level ladder-type configuration with broken symmetry. The weak probe field $\Omega_p$ couples the energy levels $|1\rangle$ and $|2\rangle$, while the strong control field $\Omega_c$ drives the transition between $|2\rangle$ and $|3\rangle$. Through a nonlinear TWM process, a signal field $\Omega_s$ is generated via the transition from level $|3\rangle$ to $|1\rangle$. The parameters $\Gamma_{31}$, $\Gamma_{32}$, and $\Gamma_{21}$ denote the decay rates of the corresponding upper energy levels. The detunings $\Delta_p$, $\Delta_c$, and $\Delta_s$ represent the single-photon detunings between each optical field and its associated transition.

Let $|n\rangle$ and $\varepsilon_n$ denote the eigenstates and eigenenergies of the free Hamiltonian $\hat{H}_0$, such that $\hat{H}_0|n\rangle = \varepsilon_n|n\rangle$. The free Hamiltonian can therefore be expressed as $\hat{H}_0 = \sum_n \varepsilon_n |n\rangle\langle n|$. As illustrated in Fig. 1, we investigate a cyclic, closed-loop configuration in a three-level ladder-type system, where otherwise forbidden transitions are enabled. Under the rotating-wave and dipole approximations, the system Hamiltonian can be reformulated as:

$$\begin{aligned}\hat{H} &= -\hbar\Delta_p|2\rangle\langle 2| - \hbar(\Delta_p+\Delta_c)|3\rangle\langle 3| \\ &\quad -\hbar\left(\Omega_p e^{i\mathbf{k}_p\cdot\mathbf{r}}|2\rangle\langle 1| + \Omega_c e^{i\mathbf{k}_c\cdot\mathbf{r}}|3\rangle\langle 2| + \Omega_s e^{i\mathbf{k}_s\cdot\mathbf{r}}|3\rangle\langle 1| + \mathrm{H.c.}\right)\end{aligned}. \tag{3}$$

Here, the detunings are defined as $\Delta_p = \omega_p - \omega_{21}$ and $\Delta_c = \omega_c - \omega_{32}$, where $\omega_p$ and $\omega_c$ are the central frequencies of the probe and control fields, respectively, and $\omega_{21}$, $\omega_{32}$ are the corresponding atomic transition frequencies. $\mathbf{r}$ denotes the position vector, and $\mathbf{k}_j\,(j=p,c,s)$ represents the wavevector of the corresponding light field. The Rabi frequencies $\Omega_j\,(j=p,c,s)$ are defined as:

$$\Omega_p = -\frac{g_y\mu_B H_p}{2\hbar}\langle 2|\hat{S}_y|1\rangle,\ \Omega_c = -\frac{g_x\mu_B H_c}{2\hbar}\langle 3|\hat{S}_x|2\rangle,\ \Omega_s = -\frac{g_y\mu_B H_s}{2\hbar}\langle 3|\hat{S}_y|1\rangle. \tag{4}$$

Here, $H_j\,(j=p,c,s)$ denotes the magnetic amplitude of the incident fields. Notably, when the $z$-axis aligns with the easy magnetization axis and the $x$-axis aligns with the DC magnetic field, the symmetry of the eigenstates ensures that all matrix elements $\langle m|\hat{S}_\nu|n\rangle, (\nu = x,y;\ m,n=1,2,3)$ in Eq. (4) are non-zero. To quantify the effect of symmetry breaking, we calculate the magnetic-dipole matrix elements using the eigenstates obtained from exact diagonalization of $\hat{H}_0$. For $H_0 = 2000G$, we obtain $|\langle 2|\hat{S}_y|1\rangle| = 0.6843$, $|\langle 3|\hat{S}_x|2\rangle| = 0.9137$, and $|\langle 3|\hat{S}_y|1\rangle| = 0.7786$. In particular, $|\langle 3|\hat{S}_y|1\rangle| \neq 0$ quantitatively demonstrates the symmetry-breaking activation of the nominally forbidden $|1\rangle \leftrightarrow |3\rangle$ transition. In the high-symmetry pure-spin regime, this transition is associated with $\Delta M_s = 3$ and has zero first-order magnetic-dipole strength.

The rhombic anisotropy mixes the relevant spin components and produces the finite value found above. It is also important to note that the system Hamiltonian is formulated in the rotating frame of the optical fields, imposing a constraint on the detunings: $\Delta_s = \Delta_p + \Delta_c$. In our analysis, we assume the control field is resonant with the atomic transition, i.e., $\Delta_c = 0$.

Since magnetic molecules do not interact with each other, it is convenient to describe the system's statistical properties using a time- and position-dependent single-particle density matrix. Accordingly, the dynamics of the three-level ladder-type system can be described by the Liouville equation:

$$\frac{\partial}{\partial t}\rho = \sum_j \Gamma_j \left( L_j \rho L_j^\dagger - \frac{1}{2}\left\{ L_j^\dagger L_j, \rho \right\} \right) - \frac{i}{\hbar}\left[ H, \rho \right], \tag{5}$$

where the first term on the right-hand side accounts for dissipative processes, with $L_j$ denoting the Lindblad operator associated with the $j$-th dissipation channel, whereas the second term describes the coherent evolution.

By substituting Eq. (3) into Eq. (5), the explicit equations of motion for the three-level ladder-type system can be obtained as:

$$\frac{\partial}{\partial t}\rho_{11} = \Gamma_{31}\rho_{33} + \Gamma_{21}\rho_{22} + i\Omega_p^*\rho_{21} + i\Omega_s^*\rho_{31} - i\Omega_p\rho_{12} - i\Omega_s\rho_{13},$$

$$\frac{\partial}{\partial t}\rho_{22} = \Gamma_{32}\rho_{33} + \Gamma_{21}\rho_{22} + i\Omega_p\rho_{12} + i\Omega_c^*\rho_{32} - i\Omega_c\rho_{23} - i\Omega_p^*\rho_{21},$$

$$\frac{\partial}{\partial t}\rho_{33} = -\left(\Gamma_{31} + \Gamma_{32}\right)\rho_{33} + i\Omega_c\rho_{23} + i\Omega_s\rho_{13} - i\Omega_c^*\rho_{32} - i\Omega_s^*\rho_{31},$$

$$\frac{\partial}{\partial t}\rho_{21} = -\left( \frac{\Gamma_{21}}{2} - i\Delta_p \right)\rho_{21} + i\Omega_p\left(\rho_{11} - \rho_{22}\right) + i\Omega_c^*\rho_{31} - i\Omega_s\rho_{23},$$

$$\frac{\partial}{\partial t}\rho_{31} = -\left[ \frac{\Gamma_{31} + \Gamma_{32}}{2} - i\left(\Delta_p + \Delta_c\right) \right]\rho_{31} + i\Omega_s\left(\rho_{11} - \rho_{33}\right) + i\Omega_c\rho_{21} - i\Omega_p\rho_{32},$$

$$\frac{\partial}{\partial t}\rho_{32} = -\left[ \frac{\Gamma_{31} + \Gamma_{32} + \Gamma_{21}}{2} - i\Delta_c \right]\rho_{32} + i\Omega_c\left(\rho_{22} - \rho_{33}\right) + i\Omega_s\rho_{12} - i\Omega_p^*\rho_{31}. \tag{6}$$

Since the system is closed, the total population is conserved, i.e., $\rho_{11} + \rho_{22} + \rho_{33} = 1$. The Hermitian property of the density matrix also requires that $\rho_{ij} = \rho_{ji}^*$. The parameters $\Gamma_{21}$, $\Gamma_{31}$, and $\Gamma_{32}$ denote the decay rates from the upper energy levels to their respective lower levels, whereas $\Gamma_3$ represents the total decay rate of the upper state $|3\rangle$, satisfying the relation $\Gamma_{31} + \Gamma_{32} = \Gamma_3$. For simplicity, we assume $\Gamma_{31} = \Gamma_{32} = \Gamma_3/2$.

For the selected three-level system, we assume that the probe field $\Omega_p$ and the signal field $\Omega_s$ are both much weaker than the control field $\Omega_c$, i.e., $\Omega_p, \Omega_s \ll \Omega_c$. Under this assumption, $\Omega_c$ is treated as constant in both time $t$ and spatial coordinate $z$. The system is initially assumed to be in the ground state $|1\rangle$, such that $\rho_{11}^{(0)} = 1$, $\rho_{22}^{(0)} = \rho_{33}^{(0)} = 0$, and $\rho_{ij}^{(0)} = 0$ for $i \neq j$. Applying a perturbative expansion to Eq. (6), $\rho_{ij} = \rho_{ij}^{(0)} + \Omega_p\rho_{ij}^{(1)} + \Omega_s\rho_{ij}^{(1)} + \ldots$, the steady-state analytical expressions for $\rho_{21}$ and $\rho_{31}$ are given by:

$$\rho_{21}=\frac{2i\Gamma_3\Omega_p+4\Delta_p\Omega_p-4\Omega_s\Omega_c^*}{\left(\Gamma_{21}-2i\Delta_p\right)\left(\Gamma_3-2i\Delta_p\right)+4\left|\Omega_c\right|^2}, \tag{7}$$

$$\rho_{31}=\frac{2i\Gamma_{21}\Omega_s+4\Delta_p\Omega_s-4\Omega_p\Omega_c}{\left(\Gamma_{21}-2i\Delta_p\right)\left(\Gamma_3-2i\Delta_p\right)+4\left|\Omega_c\right|^2}, \tag{8}$$

here, we assume that the strong control field $\Omega_c$ is on resonance with the atomic transition, i.e., $\Delta_c=0$.

The propagation of the probe field $\Omega_p$ and the signal field $\Omega_s$ through the medium can be described by the Maxwell-Bloch equations:

$$\left(\frac{1}{c}\frac{\partial}{\partial t}+\frac{\partial}{\partial z}\right)\Omega_p=i\frac{\Gamma_{21}\alpha_p}{2L}\rho_{21}, \tag{9}$$

$$\left(\frac{1}{c}\frac{\partial}{\partial t}+\frac{\partial}{\partial z}\right)\Omega_s=i\frac{\Gamma_{31}\alpha_s}{2L}\rho_{31}, \tag{10}$$

where $\alpha_j\left(j=p,s\right)$ denotes the optical density associated with each field. For simplicity, we assume $\alpha_p=\alpha_s=\alpha$, and $L$ denotes the length of the medium. Note that the phase-matching condition is satisfied, i.e., $\mathbf{k}_s=\mathbf{k}_p+\mathbf{k}_c$.

By substituting Eqs. (7) and (8) into Eqs. (9) and (10), and applying the boundary conditions $\Omega_p\left(z=0\right)=\Omega_p\left(0\right)$ and $\Omega_s\left(z=0\right)=0$, the steady-state solutions for the probe field $\Omega_p$ and the signal field $\Omega_s$ propagating through the medium are given by:

$$\Omega_p\left(z\right)=\frac{\Omega_p\left(0\right)}{2X}e^{-\frac{z\alpha\left(3\Gamma_{21}\Gamma_3-2i\left(2\Gamma_{21}+\Gamma_3\right)\Delta_p+X\right)}{4LY}}\left[\left(\Gamma_{21}\Gamma_3-4i\Gamma_{21}\Delta_p+2i\Gamma_3\Delta_p\right)\left(1-e^{\frac{z\alpha X}{2LY}}\right)+X\left(1+e^{\frac{z\alpha X}{2LY}}\right)\right], \tag{11}$$

$$\Omega_s\left(z\right)=-\frac{\Omega_p\left(0\right)\Omega_c}{X}\left[2i\Gamma_3e^{-\frac{z\alpha\left(3\Gamma_{21}\Gamma_3-2i\left(2\Gamma_{21}+\Gamma_3\right)\Delta_p+X\right)}{4LY}}\left(-1+e^{\frac{z\alpha X}{2LY}}\right)\right], \tag{12}$$

where $X=\sqrt{\left[\Gamma_{21}\left(\Gamma_3-4i\Delta_p\right)+2i\Gamma_3\Delta_p\right]^2-32\Gamma_{21}\Gamma_3\left|\Omega_c\right|^2}$ and $Y=\left(\Gamma_{21}-2i\Delta_p\right)\left(\Gamma_3-2i\Delta_p\right)+4\left|\Omega_c\right|^2$. In Appendix A, we compare the analytical results with numerical simulations obtained from the full density-matrix equations, confirming the validity of the perturbative approximation adopted throughout this work.

Before analyzing the theoretical results, it is essential to introduce the commonly used Laguerre-Gaussian (LG) vortex beams. The complex amplitude of an LG beam in cylindrical coordinates is given by [47,64]:

$$\Omega\left(r,\varphi\right)=\Omega_0\frac{1}{\sqrt{|l|!}}\left(\frac{\sqrt{2}r}{w_0}\right)^{|l|}L_p^{|l|}\left(\frac{2r^2}{w_0^2}\right)e^{-\frac{r^2}{w_0^2}}e^{il\varphi}, \tag{13}$$

where, $\Omega_0$, $r$, $w_0$, $l$, and $p$ denote the field amplitude, radial coordinate, beam waist, topological charge (azimuthal index), and radial index, respectively. $L_p^{|l|}$ is the associated Laguerre polynomial, which takes the form:

$$L_p^{|l|}(x) = \frac{e^x x^{-|l|}}{p!} \frac{d^p}{dx^p}\left(x^{|l|+p} e^{-x}\right), \tag{14}$$

where $x = 2r^2/w_0^2$ characterizes the radial dependence of the LG beam for different radial indices. When $l \neq 0$, the LG beam carries OAM along the propagation axis.

## 3. Results and Discussion

### A. Transfer of vortex light

Based on the preceding theoretical analysis, we now turn to the discussion of vortex light transfer. As shown in Eqs. (11)-(14), the vortex characteristics of the generated signal field are determined by those of the incident probe field $\Omega_p(0)$ and control field $\Omega_c$ at the entrance of the medium ($z = 0$) [see Eq. (12)]. Since either the probe field $\Omega_p$ or the control field $\Omega_c$ can carry OAM, we consider the following two cases for clarity: **A.1.** One of the beams in the input field is a vortex beam (alternatively, the probe field can be treated as a vortex beam). **A.2.** Both the probe and control fields carry OAM.

### A.1 Only the probe field is vortex light

For simplicity, we assume that only the probe field $\Omega_p(0)$ is a LG beam at the entrance of the medium ($z = 0$), as described by Eq. (13). According to Eqs. (11) and (12), the generated signal field $\Omega_s(z)$ inherits the same vortex characteristics as the incident probe field $\Omega_p(0)$, indicating a transfer of OAM from the probe to the signal field. Fig. 2(a) illustrates the variation in intensity and phase distributions of the generated signal field $\Omega_s$ at the output of the medium ($z = L$), as a function of the detuning $\Delta_p$, for an input probe field $\Omega_p(0)$ in the vortex mode $LG_2^3$ (i.e., $p_p = 2$, $l_p = 3$), with a control field $\Omega_c = 0.1\Gamma_3$. As shown in the figure, when $\Delta_p = 0$, the vortex characteristics of the probe field $\Omega_p(0)$ are effectively transferred to the generated signal field. Specifically, the intensity distribution exhibits $p_p + 1$ concentric rings, with a central dark core corresponding to a phase singularity, around which the phase winds by $2\pi l_p$. Ideally, the phase distribution should exhibit $p_p + 1$ distinct regions. However, additional parameter terms in Eq. (12) influence the phase factor of the generated field, resulting in deviations from this ideal $p_p + 1$ region structure. As $\Delta_p$ increases from -0.05$\Gamma_3$ to 0.05$\Gamma_3$, both the intensity and phase of the generated signal field are modulated accordingly. In particular, as $\Delta_p$ increases from 0 to 0.05$\Gamma_3$, the signal field intensity becomes increasingly concentrated in the inner ring, while the outer ring intensity diminishes. This behavior arises because the detuning modifies the dispersive response of the medium, thereby changing the phase accumulated by the propagating field. The resulting refractive-index variation leads to a spatial redistribution of the intensity of the generated signal field. When $\Delta_p < 0$, the resulting phase distortions exhibit the opposite trend. Fig. 2(b) shows that, for fixed detuning $\Delta_p = 0$, the intensity and phase of the generated signal field $\Omega_s$ are also modulated as the control field strength $\Omega_c$ increases from 0.1$\Gamma_3$ to 2$\Gamma_3$. In this case, the outer ring intensity gradually diminishes, while the inner ring intensity becomes increasingly prominent. This behavior is attributed to additional parameter terms in Eq. (12), which allow variations in $\Omega_c$ to indirectly influence the phase evolution of the signal field. Although the overall phase profile remains undistorted, it becomes segmented into multiple regions, thereby affecting the spatial redistribution of the signal field intensity.

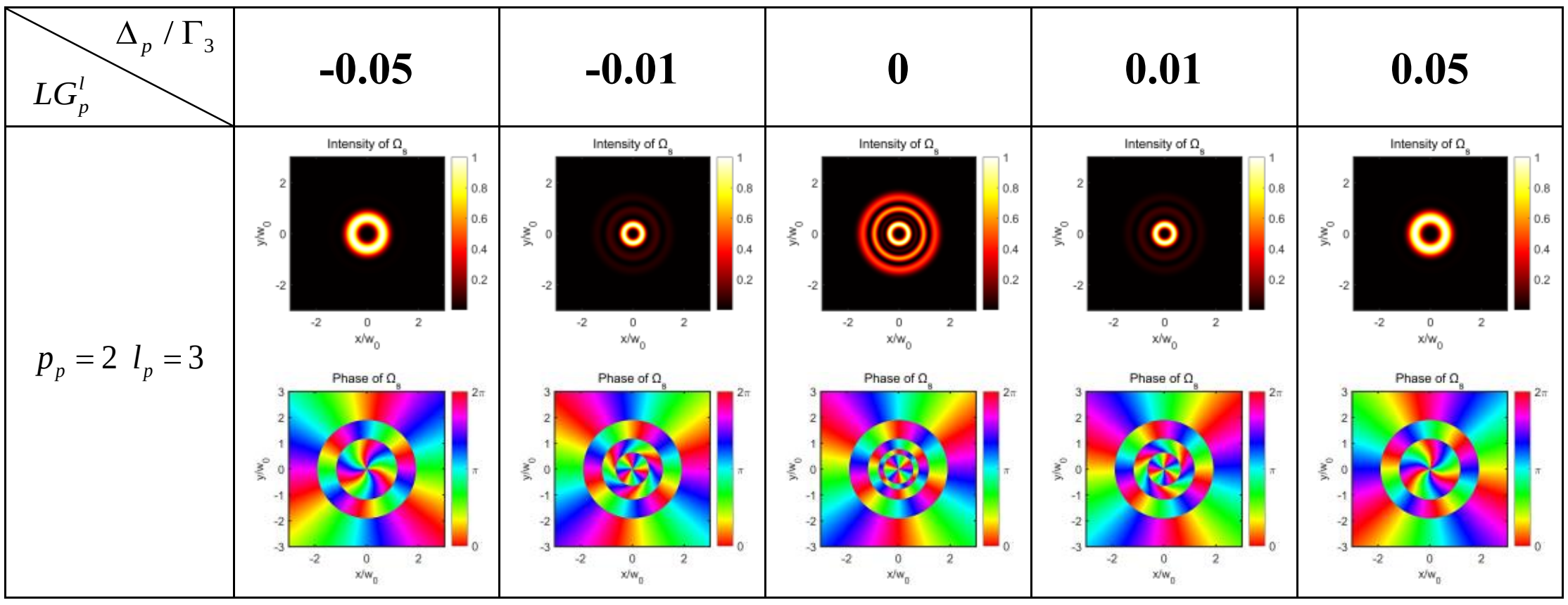

| $\Delta_p/\Gamma_3$ \ $LG_p^l$ | **-0.05** | **-0.01** | **0** | **0.01** | **0.05** |
|---|---|---|---|---|---|
| $p_p=2$ $l_p=3$ | | | | | |

**Fig. 2(a).** When the probe field $\Omega_p(0)$ is prepared in the $LG_2^3$ vortex mode, the intensity and phase of the generated signal field $\Omega_s$ vary as a function of the detuning $\Delta_p$ ( $\Delta_p=-0.05\Gamma_3,-0.01\Gamma_3,0,0.01\Gamma_3,0.05\Gamma_3$ ). The remaining parameters are set as follows: $\Omega_{p0}=0.01\Gamma_3$, $\Omega_c=0.1\Gamma_3$, $\alpha=30$, $\Gamma_{21}=0.05\Gamma_3$, and $z=L$.

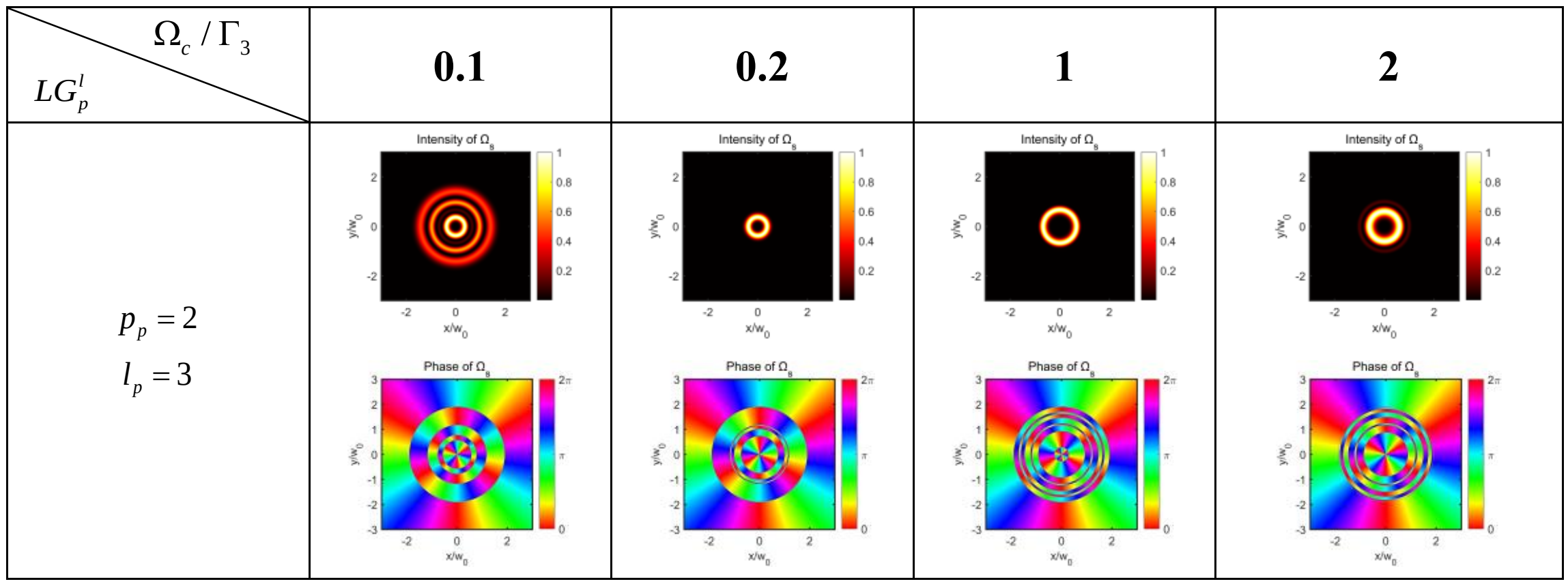

| $\Omega_c/\Gamma_3$ \ $LG_p^l$ | **0.1** | **0.2** | **1** | **2** |
|---|---|---|---|---|
| $p_p=2$ $l_p=3$ | | | | |

**Fig. 2(b).** When the probe field $\Omega_p(0)$ is prepared in the $LG_2^3$ vortex mode, the intensity and phase of the generated signal field $\Omega_s$ vary with the strength of the control field $\Omega_c$ ($\Omega_c=0.1\Gamma_3,0.2\Gamma_3,1\Gamma_3,2\Gamma_3$). The remaining parameters are set as follows: $\Omega_{p0}=0.01\Gamma_3$, $\Delta_p=0$, $\alpha=30$, $\Gamma_{21}=0.05\Gamma_3$, and $z=L$.

### A.2. Both the probe and control fields are vortex beams

We next consider a more general scenario in which both the probe field $\Omega_p(0)$ and the control field $\Omega_c$ at the entrance of the medium ($z=0$) are vortex beams. Here, we focus exclusively on the case with radial index $p=0$. According to Eqs. (11)-(14), the OAM carried by the generated signal field $\Omega_s(z)$ corresponds to the sum of the OAMs of the incident probe field $\Omega_p(0)$ and the control field $\Omega_c$, such that the topological charges satisfy $l_s=l_p+l_c$. Fig. 3(a) illustrates the intensity and phase distributions of the generated signal field $\Omega_s$ as a function of the detuning $\Delta_p$, where the input probe field $\Omega_p(0)$ and the control field $\Omega_c$ are prepared in the $LG_0^3$ mode (i.e., $p_p=0$, $l_p=3$) and the $LG_0^2$ mode (i,e., $p_c=0$, $l_c=2$), respectively. As shown in the figure, when $\Delta_p=0$, the generated signal field exhibits a circular intensity pattern, which ideally forms a bright ring. The appearance of the central dark ring arises from the combined influence of other parameter terms in Eq. (12). As the detuning $\Delta_p$ increases from -0.05$\Gamma_3$ to 0.05$\Gamma_3$, both the amplitude and phase of the generated signal field are modulated. Specifically, as $\Delta_p$ increases from 0 to 0.05$\Gamma_3$, the intensity of the generated signal field gradually redistributed toward the inner ring. This behavior arises from the detuning-induced variation in the dispersion response of the medium, which leads to phase distortion. For $\Delta_p<0$, the direction of phase distortion is reversed. Fig. 3(b) illustrates the effect of varying the control field intensity $\Omega_{c0}$ on the intensity and phase of the generated signal field $\Omega_s$ under zero detuning ($\Delta_p=0$). Specifically, as the control field modulates the signal field, the outer ring intensity gradually decreases, whereas the inner ring intensity becomes increasingly pronounced, accompanied by the fragmentation of the phase profile into multiple distinct domains.

| $\Delta_p/\Gamma_3$ \ $LG_{p=0}^l$ | **-0.05** | **-0.01** | **0** | **0.01** | **0.05** |
|---|---|---|---|---|---|
| $l_p=3$<br>$l_c=2$ | 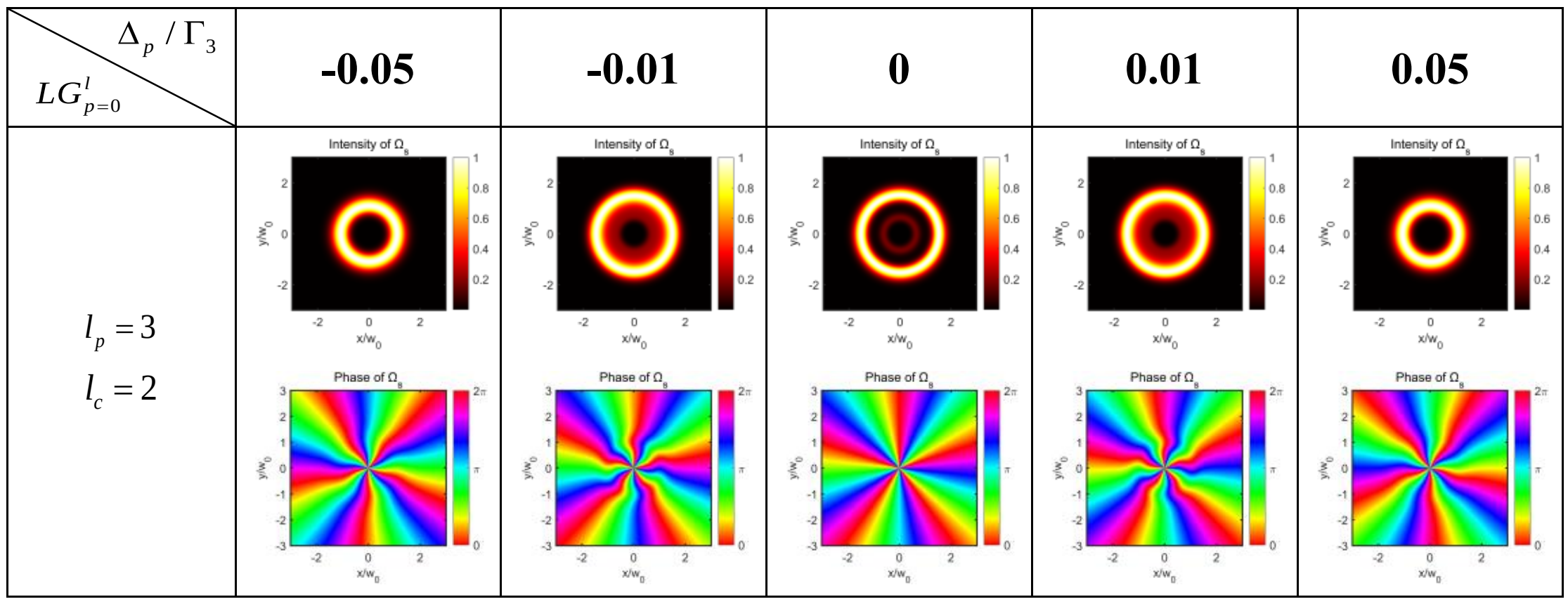 | | | | |

**Fig. 3(a).** When the probe field $\Omega_p(0)$ and the control field $\Omega_c$ are prepared in the vortex modes $LG_0^3$ and $LG_0^2$, respectively, the intensity and phase of the generated signal field $\Omega_s$ vary with the detuning $\Delta_p$ ($\Delta_p=-0.05\Gamma_3, -0.01\Gamma_3, 0, 0.01\Gamma_3, 0.05\Gamma_3$). The remaining parameters are set as follows: $\Omega_{p0}=0.01\Gamma_3$, $\Omega_{c0}=0.1\Gamma_3$, $\alpha=30$, $\Gamma_{21}=0.05\Gamma_3$, and $z=L$.

| $\Omega_{c0}/\Gamma_3$ \ $LG_{p=0}^l$ | **0.1** | **0.2** | **1** | **2** |
|---|---|---|---|---|
| $l_p=3$<br>$l_c=2$ | | | | |

**Fig. 3(b).** When the probe field $\Omega_p(0)$ and the control field $\Omega_c$ are prepared in the vortex modes

$LG_0^3$ and $LG_0^2$, respectively, the intensity and phase of the generated signal field $\Omega_s$ vary with the control field $\Omega_{c0}$ ($\Omega_{c0} = 0.1\Gamma_3, 0.2\Gamma_3, 1\Gamma_3, 2\Gamma_3$). The remaining parameters are set as follows: $\Omega_{p0} = 0.01\Gamma_3$, $\Delta_p = 0$, $\alpha = 30$, $\Gamma_{21} = 0.05\Gamma_3$, and $z = L$.

To quantitatively characterize the transverse intensity and phase distributions of the generated signal field, we express the signal field as

$$\Omega_s(L,r) = T_s(r)\Omega_p(0,r), \tag{15}$$

where

$$T_s(r) = \frac{\Omega_s(L,r)}{\Omega_p(0,r)}, \tag{16}$$

is the local complex transfer function obtained from Eq. (12). The corresponding intensity and phase of the signal field can be expressed as

$$I_s(r) = |\Omega_p(0,r)|^2 |T_s(r)|^2, \tag{17}$$

and

$$\delta\Phi_{med}(r) = \arg\left[T_s(r)\right], \tag{18}$$

respectively. The radial phase gradient is evaluated as

$$g_r(r) = \frac{\partial\delta\Phi_{med}(r)}{\partial(r/w_0)} = \frac{\mathrm{Im}\left[T_s^*(r)\partial T_s(r)/\partial(r/w_0)\right]}{|T_s(r)|^2}. \tag{19}$$

As shown in Fig. 2(a), the radial intensity, phase, and phase gradient all vary systematically with the probe detuning $\Delta_p$. At resonance, the continuous radial phase gradient is approximately zero away from zero-intensity radii. For $|\Delta_p| = 0.01\Gamma_3$, the radial phase excursion reaches approximately 0.50 rad, and the maximum phase-gradient magnitude in the appreciable-intensity region is approximately 1.70. When the detuning is increased to $|\Delta_p| = 0.05\Gamma_3$, these values increase to approximately 2.13 rad and 2.43, respectively. Reversing the sign of the detuning reverses the direction of the radial phase variation, consistent with the change in the dispersive response. In particular, $\delta\Phi_{med}(r,\Delta_p) \approx -\delta\Phi_{med}(r,-\Delta_p)$ and $g_r(r,\Delta_p) \approx -g_r(r,-\Delta_p)$. By comparison, the magnitude of the transfer function is approximately symmetric with respect to the sign of the detuning, so that $I_s(r,\Delta_p) \approx I_s(r,-\Delta_p)$. This explains why opposite detunings can produce similar radial intensity distributions while inducing radial phase with opposite directions. The intensity redistribution and radial phase modulation are two associated consequences of the same complex transfer function. Specifically, $|T_s(r)|$ determines the position-dependent conversion, gain, and absorption, whereas $\arg\left[T_s(r)\right]$ describes the corresponding dispersive phase accumulation. The observed redistribution refers to a change in the relative intensities and positions of the output rings caused by the radial dependence of $|T_s(r)|$. The simultaneous variation of the radial phase provides evidence that the amplitude reshaping and phase modulation originate from the same complex optical response. For Fig. 2(b), the probe detuning is fixed at $\Delta_p = 0$, while the control field is varied from $\Omega_c = 0.1\Gamma_3$ to $\Omega_c = 2\Gamma_3$. Owing to the presence of the control field in both the nonlinear coupling term and the complex response in Eq. (12), $T_s(r)$ depends nonlinearly on $|\Omega_c(r)|$. Increasing $\Omega_c$ therefore changes the radial positions at which the local conversion efficiency is maximized. The incident $LG_2^3$ probe contains several radial rings separated by radial nodes. Because the Gaussian control field decreases monotonically with $r$, the inner and outer rings experience different local control strengths. Changing $\Omega_c$ consequently changes the value of $|T_s(r)|$ differently in these radial regions, leading to the enhancement or suppression of individual rings and to shifts of the dominant intensity maximum in Fig. 2(b). When $\Delta_p = 0$, the medium-induced phase is radially constant, apart from phase-branch changes at zero-intensity positions; consequently, and the continuous radial phase gradient is approximately zero. For the vortex-control configuration in Figs. 3(a) and 3(b), the known azimuthal phase factor of the control field is first separated from the transfer function. The remaining radial transfer function is then defined and analyzed in the same way as $T_s(r)$ above. In Fig. 3(a), the probe and control fields are the $LG_0^3$ and $LG_0^2$ modes, respectively. Unlike the Gaussian control field in Fig. 2, this vortex control vanishes at the beam center, increases with radius, reaches its maximum at $r/w_0 = 1$ and subsequently decreases in the outer radial region. Therefore, different radial positions within the same probe beam experience substantially different control-field strengths and may correspond to different local coupling regimes. The radial phase analysis corresponding to Fig. 3(a) shows that, for $|\Delta_p| = 0.01\Gamma_3$, the maximum medium-induced phase variation is approximately 1.66 rad, and the maximum magnitude of the radial phase gradient is approximately 5.22. For $|\Delta_p| = 0.05\Gamma_3$ the corresponding values are approximately 0.83 rad and 1.65. Unlike the Gaussian-control case, these quantities do not vary monotonically with $|\Delta_p|$. This behavior results from the nonmonotonic radial variation of the vortex-control amplitude. As $r/w_0$ increases, the local control strength first increases and then decreases, causing different radial regions to sample different parts of the complex response. As in Fig. 3(a), changing the sign of the detuning reverses the sign of the dispersive phase and its radial gradient but leaves the intensity profile approximately unchanged. In Fig. 3(b), the probe detuning remains fixed at $\Delta_p = 0$, while the control field is varied from $0.1\Gamma_3$ to $2\Gamma_3$. Increasing $\Omega_{c0}$ changes the radial positions at which the control field provides the most efficient nonlinear conversion. It therefore modifies $|T_s(r)|$, shifts the dominant signal ring, and can enhance or suppress secondary radial rings, as observed in Fig. 3(b). At exact resonance, the medium-induced radial phase is approximately constant in every finite-intensity region, and its continuous radial gradient is approximately zero.

It is worth noting that in both scenarios described above, although variations in the detuning $\Delta_p$ or the control field $\Omega_c$ may lead to deviations in the radial index $p$ of the vortex beams, the number of phase cycles per azimuthal region remains equal to $|l_p|$. This observation reflects the topological stability of the vortex beam modes [36]. This can be understood by separating the radial amplitude, after which the generated signal field can be written as

$$\Omega_s(L,r,\varphi) = B_s(r)e^{i(l_p+l_c)\varphi}, \tag{20}$$

where $B_s(r)$ contains the radial amplitude and the medium-induced radial phase. The topological charge evaluated along a closed loop is

$$l_s = \frac{1}{2\pi}\oint \nabla_\perp \arg\left[\Omega_s(L,r,\varphi)\right]\cdot d\ell = l_p + l_c. \tag{21}$$

Radial phase variations, including possible $\pi$-phase jumps across radial nodes, do not change this azimuthal winding number.

**B. Comparison of EIT and ATS in the gain characteristics of the vortex signal field**

Distinguishing between EIT and ATS in a three-level system is a subtle challenge, as both mechanisms can produce a transparency window in the absorption spectrum. However, the underlying physical mechanisms are fundamentally different. EIT arises from Fano interference between two transition pathways in a three-level system [57], whereas ATS results from the dynamic splitting of the absorption spectrum into two peaks under strong coupling fields, a phenomenon described by the AC-Stark effect [60]. This distinction has been extensively investigated and discussed in the literature [65-70]. Following Refs. [66], a threshold criterion of $\frac{1}{2}|\Gamma_3 - \Gamma_{21}|$ can be used to differentiate between the EIT and ATS regimes. When the control field $|\Omega_c| < \frac{1}{2}|\Gamma_3 - \Gamma_{21}|$, the system is considered to exhibit EIT; whereas for $|\Omega_c| > \frac{1}{2}|\Gamma_3 - \Gamma_{21}|$, ATS is expected to dominate.

We next compare the influence of EIT and ATS on the gain of the vortex signal field in the TWM process, based on a three-level ladder-type system. For clarity, we assume $\Gamma_{21} = 0.05\Gamma_3$, such that the EIT regime corresponds to $|\Omega_c| < 0.475\Gamma_3$, while the ATS regime corresponds to $|\Omega_c| > 0.475\Gamma_3$. We first examine the variation in gain, expressed as the relative intensity $|\Omega_s/\Omega_p(0)|^2$, of the generated signal field when the control field $\Omega_c$ is a conventional Gaussian beam, under the two distinct regimes. Figs. 4(a) and 4(b) present the relative intensities $|\Omega_{\mathrm{p,s}}/\Omega_{\mathrm{p}}(0)|^2$ of the probe and signal fields as functions of the propagation distance $z$, under the EIT condition ($\Omega_c = 0.2\Gamma_3$) and ATS condition ($\Omega_c = 2\Gamma_3$), respectively, for various detuning values $\Delta_p$. As shown in the figures, the gain (i.e., relative intensity) of the generated signal field decreases with increasing detuning $\Delta_p$. The gain reaches its maximum at $\Delta_p = 0$, as the optical field is resonant with the medium. A comparison of Figs. 4(a) and 4(b) reveals that the ATS regime yields a stronger signal field gain than the EIT regime under identical parameters. This observation contrasts with the conventional view that the EIT mechanism enhances nonlinear processes [17,58]. It is worth noting that the gain of the signal field, expressed as $|\Omega_s/\Omega_{\mathrm{p}}(0)|^2$, can exceed unity due to the action of a strong control field $\Omega_c$, which pumps population from level $|2\rangle$ to the upper level $|3\rangle$, thereby supplying additional energy. Consequently, although $|\Omega_s/\Omega_{\mathrm{p}}(0)|^2$ has been used in earlier studies [50,51,61] to characterize the conversion efficiency of the signal field, its interpretation may depend on the specific interaction configuration. For interaction systems involving only the probe field $\Omega_{\mathrm{p}}$ and the signal field $\Omega_s$, $|\Omega_s/\Omega_{\mathrm{p}}(0)|^2$ can reasonably serve as a measure of conversion efficiency [45]. In multi-beam interaction systems, however, this quantity should be interpreted with some care. For example, in a dual $\Lambda$-type FWM system, $|\Omega_s/\Omega_{\mathrm{p}}(0)|^2$ corresponds to the conversion efficiency only when the two strong control fields are equal, such that absorption and gain are perfectly balanced [50-52]. A more rigorous and general definition is given in Ref. [71]. Here, we interpret $\left|\Omega_s/\Omega_p(0)\right|^2$ as the gain of the generated signal field, rather than as the conversion efficiency. A detailed distinction between the gain of the signal field and the conversion efficiency is provided in Appendix B. Thus, the TWM process not only enables frequency conversion of vortex beams but also facilitates intensity amplification through the ATS mechanism.

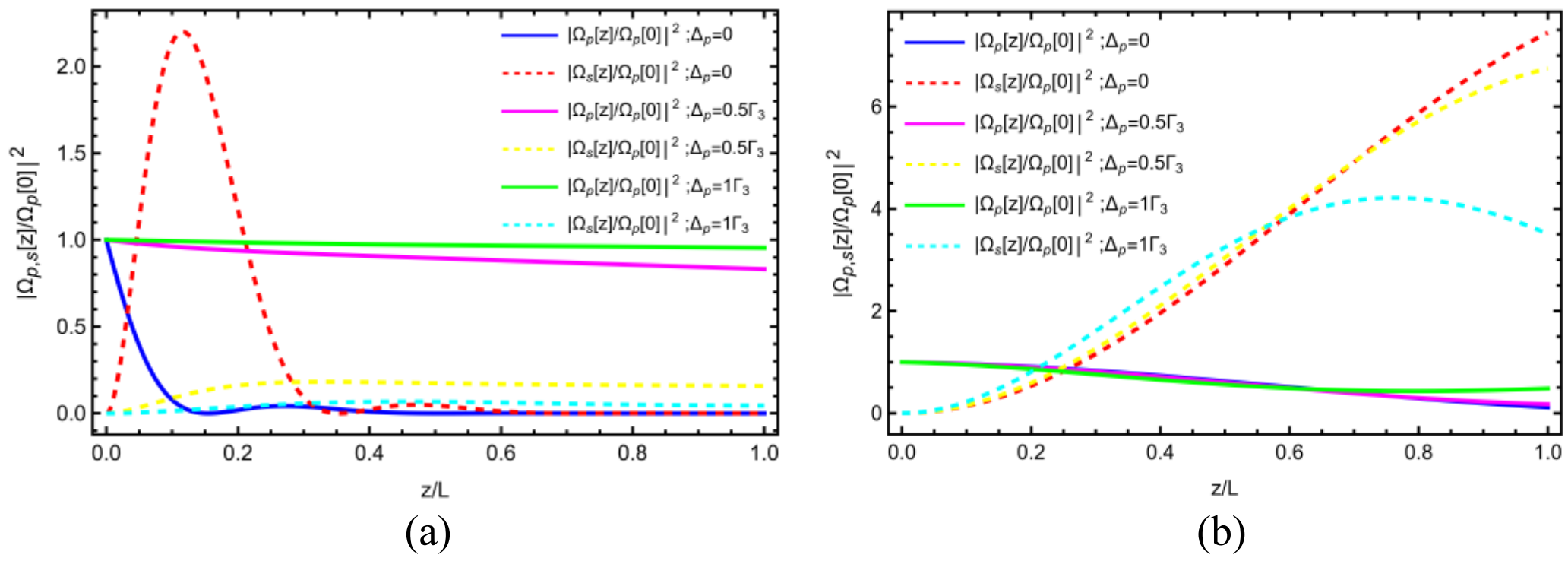

**Fig. 4.** Relative intensities $\left|\Omega_{p,s}/\Omega_p(0)\right|^2$ of the probe and signal fields as functions of the normalized propagation distance $z/L$ under different detuning values $\Delta_p$, for both EIT and ATS regimes. (a) In the EIT regime ($\Omega_c = 0.2\Gamma_3$), the relative intensities of the probe and signal fields are plotted as functions of the normalized distance $z/L$ for various detuning values $\Delta_p$. (b) In the ATS regime ($\Omega_c = 2\Gamma_3$), the relative intensities of the probe and signal fields are shown as functions of the normalized distance $z/L$ under different $\Delta_p$. The remaining parameters for both panels are: $\Delta_p = 0, 0.5\Gamma_3, 1\Gamma_3$, $\alpha = 30$, $\Omega_{p0} = 0.01\Gamma_3$, and $\Gamma_{21} = 0.05\Gamma_3$.

We next examine the evolution of the relative intensities $\left|\Omega_{p,s}/\Omega_p(0)\right|^2$ of the probe and signal fields as functions of the normalized propagation distance $z/L$, under varying control field strengths $\Omega_c$ with fixed detuning $\Delta_p = 0$. As shown in Fig. 5, the gain (i.e., relative intensity) of the generated signal field increases with increasing control field strength $\Omega_c$. This further confirms the advantage of the ATS regime over EIT in enhancing the gain of the signal field.

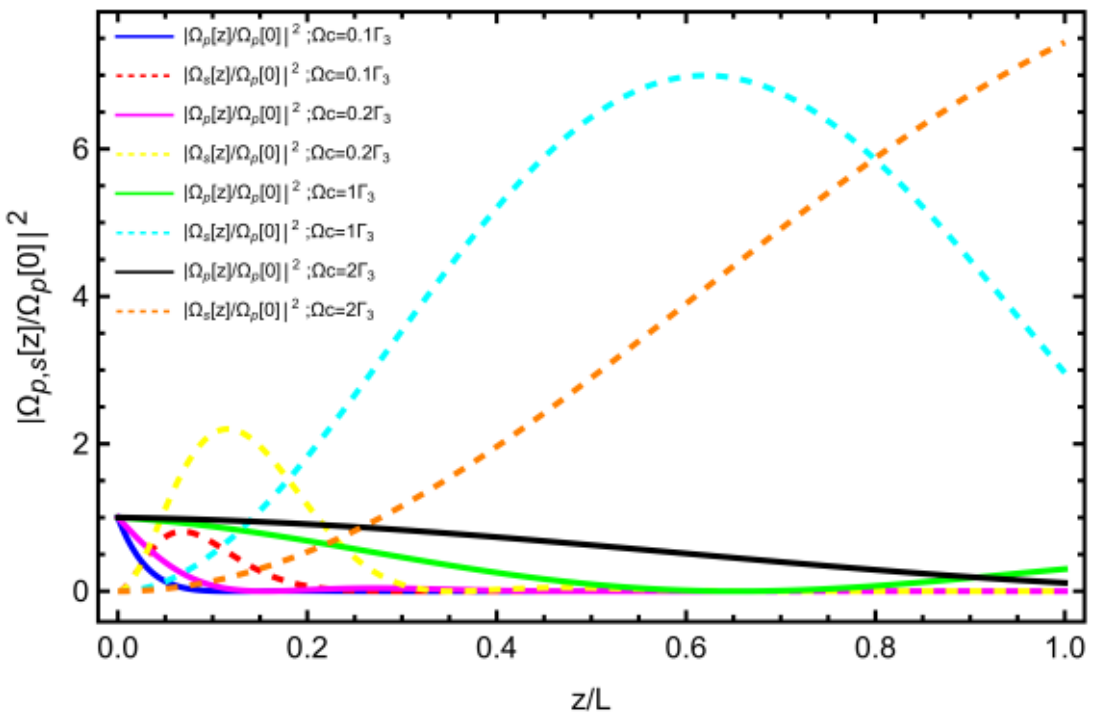

**Fig. 5.** Relative intensities $\left|\Omega_{p,s}/\Omega_p(0)\right|^2$ of the probe and signal fields as functions of the normalized propagation distance $z/L$ in the medium, under different control field strengths $\Omega_c$. The remaining parameters are: $\Delta_p = 0$, $\alpha = 30$, $\Omega_{p0} = 0.01\Gamma_3$, and $\Gamma_{21} = 0.05\Gamma_3$.

Secondly, when the control field $\Omega_c$ is a vortex beam, its intensity distribution becomes dependent on the transverse radial coordinate $r$, necessitating an analysis of how this spatial variation influences the gain (relative intensity) $\left|\Omega_s/\Omega_p(0)\right|^2$ of the signal field. Figs. 6(a), 6(c), and 6(e) show the intensity distributions of the control field $\Omega_c$ with respect to the transverse radius $r$, for different $LG_p^l$ mode configurations. As shown in the figures, the intensity profile of $|\Omega_c|$ varies significantly with $r$ when the radial index $p$ and topological charge $l$ are changed, leading to the coexistence of EIT and ATS effects at different radial positions across the beam cross section. Figs. 6(b), 6(d), and 6(f) illustrate the resulting signal field gain $|\Omega_s/\Omega_\mathrm{p}(0)|^2$ at the output end of the medium ( $z=L$ ), showing its variation with $r$ for each corresponding $LG_p^l$-mode control field shown in Figs. 6(a), 6(c), and 6(e). As shown in Figs. 6(b) and 6(d), when the control field $|\Omega_c|$ is relatively weak, the system may still fall within the ATS regime, yet the signal field gain at the output of the medium is not necessarily greater than that in the EIT regime. This behavior arises because the light field experiences absorption over the finite propagation length, so the competition between gain and loss reduces the net gain at the medium output. In contrast, when $|\Omega_c|$ is strong [as shown in Fig. 6(f)], gain dominates over loss, and the ATS regime yields a higher output gain than the EIT regime. In Appendix C, we compare the maximum gain of the generated signal field under the ATS and EIT mechanisms. We find that, under optimized operating conditions, the gain advantage of the ATS regime persists over a broad range of system parameters.

These results confirm that, within our system, the gain enhancement induced by the ATS mechanism is more pronounced than that from EIT. This advantage of ATS gain over EIT can be qualitatively understood as follows: a stronger $|\Omega_c|$ enhances the coupling between energy levels $|2\rangle$ and $|3\rangle$, thereby increasing the population pumping efficiency and facilitating the generation of the signal field $\Omega_s$.

We next provide a quantitative description by analyzing the analytical expression for the signal field gain. The gain advantage in the ATS regime can be understood more clearly by separating the coherent conversion and dissipative contributions in the analytical solution. A stronger control field enhances coherence transfer from the probe transition to the signal transition while simultaneously modifying the absorption experienced during propagation. For $\Delta_p=0$ and an undepleted control field, the analytical expression for the signal field gain (relative intensity), can be written as

$$G_s(\zeta)\equiv\frac{|\Omega_s(z)|^2}{|\Omega_p(0)|^2}=\frac{16\Gamma_3^2 C}{|Q|^2}\exp\left(-\frac{3\alpha A}{2D}\zeta\right)\left|\sinh\left(\frac{\alpha Q}{4D}\zeta\right)\right|^2, \tag{22}$$

where $\zeta=z/L$ denotes the dimensionless propagation distance, with $A=\Gamma_{21}\Gamma_3$, $C=|\Omega_c|^2$, $D=A+4C$, and $Q=\sqrt{A(A-32C)}$. The factor $16\Gamma_3^2 C/|Q|^2$ describes the control-assisted coupling strength, the exponential factor $\exp\left(-\frac{3\alpha A}{2D}\zeta\right)$ represents the dissipative propagation envelope, and the hyperbolic-sine term describes the coherent buildup and exchange of the probe and generated signal fields. Although Eq. (22)

appears singular at $Q=0$, this singularity is removable because $\lim_{Q\to 0}\frac{\sinh(Qx)}{Q}=x$. The corresponding finite result is

$$G_s(\zeta)=\frac{\alpha^2\Gamma_3^2 C}{D^2}\zeta^2\exp\left(-\frac{3\alpha A}{2D}\zeta\right)\qquad Q=0. \tag{23}$$

In the strong-control limit, $C\gg A$, the quantity $Q$ becomes approximately imaginary: $Q\approx i\sqrt{32AC}$. Using $\sinh(ix)=i\sin(x)$, Eq. (22) becomes

$$G_s^{(strong)}(\zeta)\approx\frac{\Gamma_3}{2\Gamma_{21}}\exp\left(-\frac{3\alpha A}{8C}\zeta\right)\sin^2\left[\alpha\zeta\sqrt{\frac{A}{8C}}\right], \tag{24}$$

The attenuation exponent is proportional to $A/C$, so increasing the control strength initially suppresses the dissipation and permits a substantially larger signal to reach the output. This behavior accounts for the pronounced gain obtained in the ATS regime. Nevertheless, Eq. (24) also shows that the gain cannot increase indefinitely. In the extremely strong-control limit, $C\to\infty$, the sine argument becomes small, and Eq. (24) reduces to

$$G_s^{(ultrastrong)}(\zeta)\approx\frac{\alpha^2\Gamma_3^2}{16|\Omega_c|^2}\zeta^2. \tag{25}$$

Thus, $G_s^{(ultrastrong)}\propto\frac{1}{|\Omega_c|^2}$ .An excessively strong control field therefore reduces the effective conversion over a fixed medium length. The gain reaches an optimum when the reduction of absorption and the coherent conversion length are properly balanced.

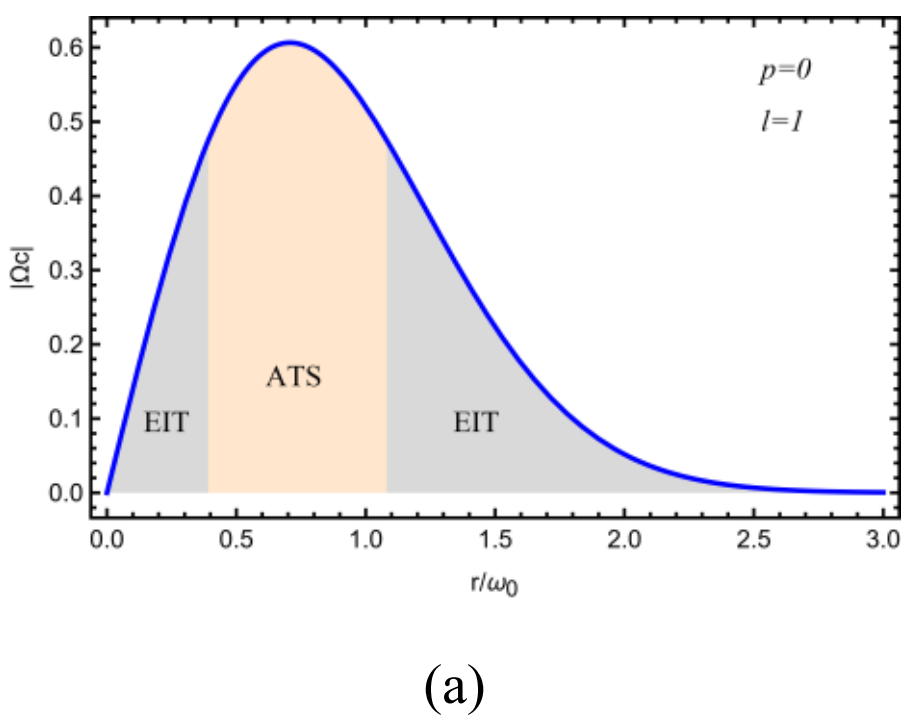

(a)

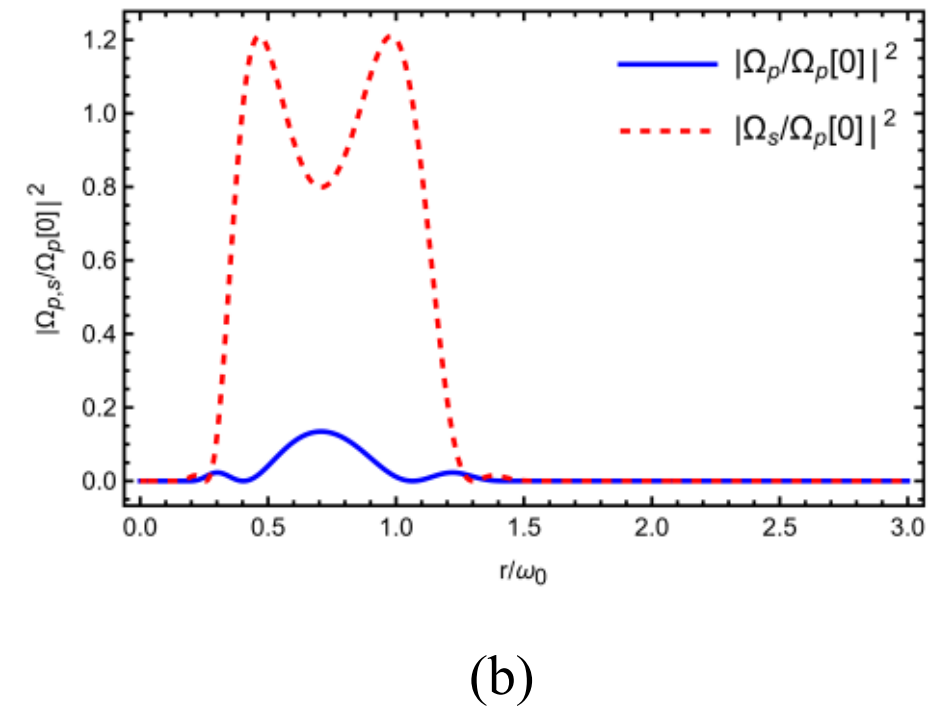

(b)

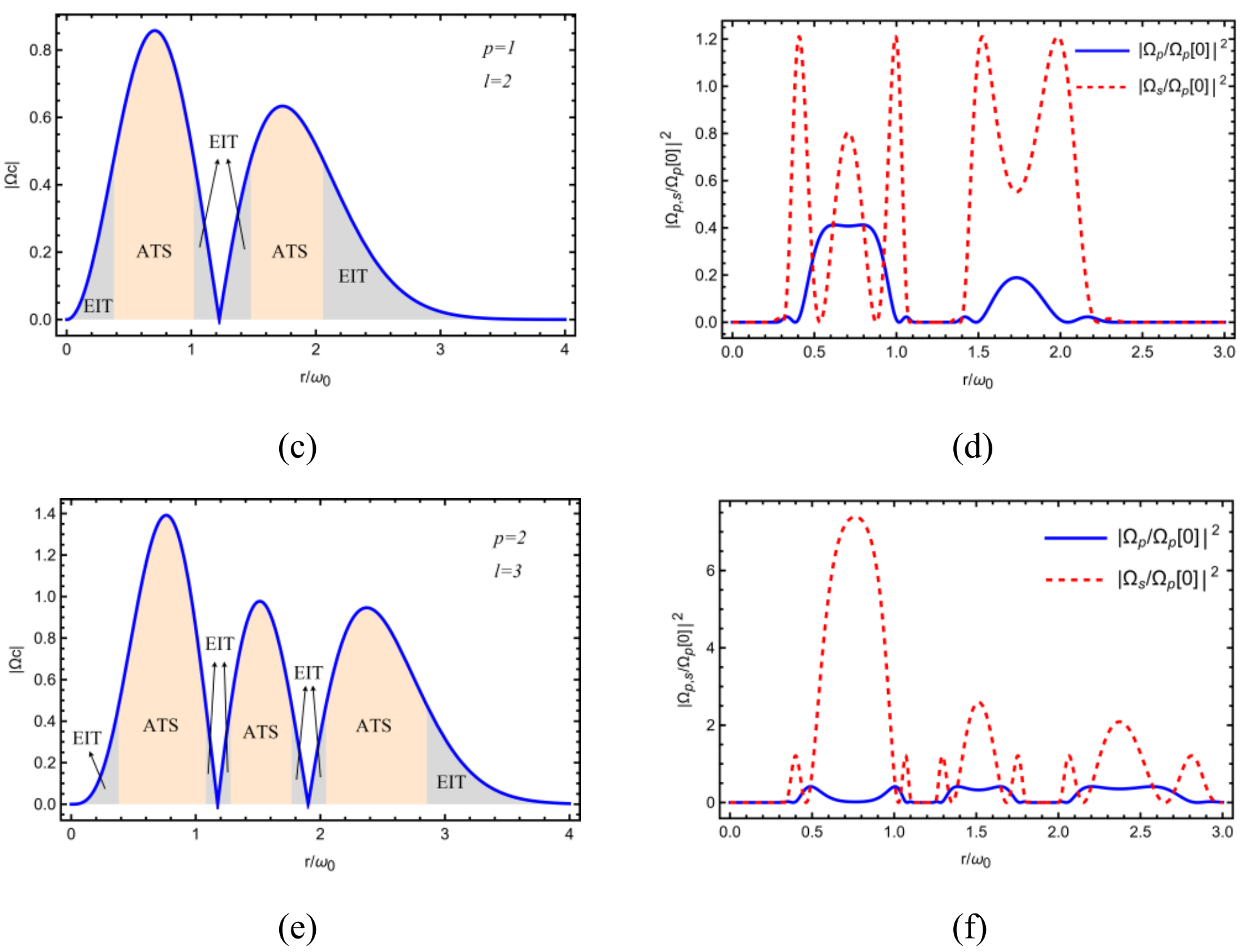

**Fig. 6.** (a), (c), and (e) show the radial intensity profiles of the control field $\Omega_c$ in different $LG_p^l$ modes. In (a), the control field is in the $LG_p^l$ mode with $p=0$, $l=1$; in (c), $p=1$, $l=2$; and in (e), $p=2$, $l=3$. (b), (d), and (f) show the corresponding signal field gain $\left|\Omega_s/\Omega_p(0)\right|^2$ at the output position $z=L$, plotted as a function of the radial coordinate $r$, for each control field mode shown in (a), (c), and (e). The remaining parameters are: $\Delta_p=0$, $\Delta_c=0$, $\alpha=30$, $\Omega_{c0}=1\Gamma_3$, $\Omega_{p0}=0.01\Gamma_3$, $\Gamma_{21}=0.05\Gamma_3$, and $z=L$.

## 4. Summary

In this work, we employed a conventional three-level ladder-type configuration in molecular magnets to enable high-gain vortex light transfer across otherwise forbidden transitions. We show that both the intensity and phase of the generated vortex signal field are governed by the detuning of the probe field and the amplitude of the control field, and that the topological charges of the vortex beams satisfy a well-defined algebraic conservation law during the transfer process. We compared the signal field gain under EIT and ATS in a nonlinear TWM process. Contrary to the conventional view that EIT is more favorable for enhancing nonlinear effects, our results demonstrate that ATS provides a larger signal-field gain than EIT over a broad parameter range in this system. We further refined the definition of signal-field conversion efficiency, specifying the conditions under which it is applicable, and highlighting its limitations in multi-field interaction systems. These findings contribute to a deeper understanding of the interaction between vortex light and matter, and, owing to the long spin coherence times of molecular magnets and their characteristic transition frequencies in the microwave regime, the results may have potential applications in quantum information storage and transmission, quantum computing, and microwave radar detection.

## Appendix A: Validity of the perturbative approximation

To verify the validity of the first-order expansion used in Eqs. (7) and (8), we directly solved the complete steady-state density-matrix equations in Eq. (6). At each propagation position, the full steady-state density matrix was obtained subject to the population-conservation condition $\rho_{11}+\rho_{22}+\rho_{33}=1$. The resulting numerical coherences $\rho_{21}$ and $\rho_{31}$ were then coupled to the Maxwell-Bloch equations in Eqs. (9) and (10), which were integrated numerically along the propagation direction. Fig. A1 compares the analytical solutions in Eqs. (11) and (12) with the full numerical solutions for the representative EIT and ATS parameters $\Omega_c=0.2\Gamma_3$ and $\Omega_c=2\Gamma_3$, respectively. The analytical curves and full numerical results nearly overlap. To quantify the agreement between them, we define the normalized root-mean-square relative error $\varepsilon_{rms}$ between the analytical solution and the full numerical solution as

$$\varepsilon_{rms}=100\sqrt{\frac{\int_0^1\left(\left|\Omega_p^{num}-\Omega_p^{ana}\right|^2+\left|\Omega_s^{num}-\Omega_s^{ana}\right|^2\right)d\left(z/L\right)}{\int_0^1\left(\left|\Omega_p^{num}\right|^2+\left|\Omega_s^{num}\right|^2\right)d\left(z/L\right)}}, \tag{A1}$$

where the superscripts "num" and "ana" denote quantities obtained from the direct numerical solution and the first-order perturbative analytical solution, respectively. At zero detuning, the calculated errors are 3.67%, 1.52%, 0.0535%, and 0.0091% for $\Omega_c=0.1\Gamma_3, 0.2\Gamma_3, 1\Gamma_3, 2\Gamma_3$, respectively. These results show that the first-order solution is accurate to within approximately 4% over the complete control-field range considered in this work. The approximation becomes increasingly accurate as the control field increases. The full numerical and analytical signal intensities differ by less than 0.1% for all detunings used in Fig. 4. The strong-control-field regime therefore lies well within, rather than outside, the range of validity of the weak-probe expansion.

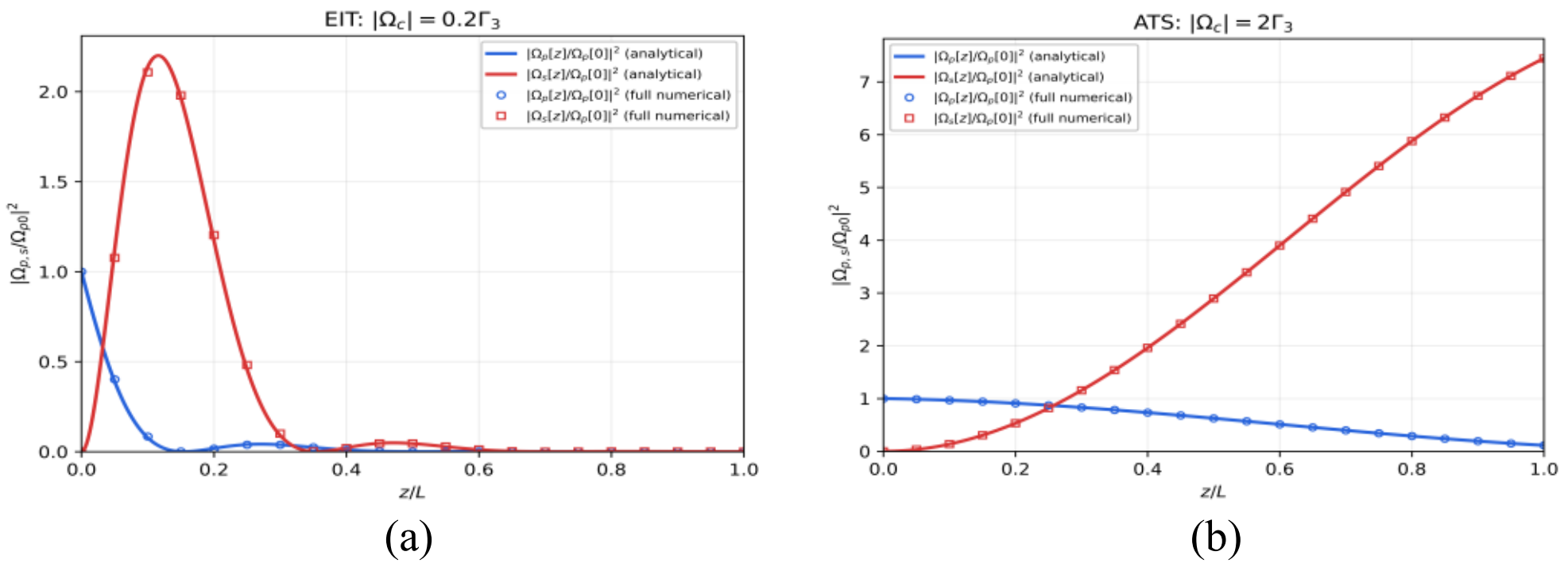

(a) (b)

**Fig. A1.** Comparison between the perturbative analytical and direct numerical solutions. Panel (a) corresponds the representative EIT condition with $\Omega_c=0.2\Gamma_3$, whereas panel (b) corresponds the ATS condition with $\Omega_c=2\Gamma_3$. The other parameters are $\Delta_p=0,$, $\alpha=30$, $\Omega_{p0}=0.01\Gamma_3$, and $\Gamma_{21}=0.05\Gamma_3$ for both panels.

## Appendix B: Comparison between signal gain and conversion efficiency

In nonlinear frequency conversion of vortex beams, several normalized quantities are often referred to as conversion efficiencies. Their physical meanings depend on the fields included in the normalization and on whether the generated field obtains energy solely from the incident probe field or from one or more input fields. In this appendix, we distinguish these definitions in detail and identify the former as a gain, whereas only the latter should be regarded as a genuine efficiency.

In studies of vortex-beam frequency conversion, the signal-field conversion efficiency is often defined as [50,51,61]

$$G_s\left(z\right)=\frac{\left|\Omega_s\left(z\right)\right|^2}{\left|\Omega_p\left(0\right)\right|^2}. \tag{B1}$$

However, this interpretation should be treated with caution, because it can depend on the specific interaction model. For an interaction system involving only the probe and signal fields, Eq. (B1) can be reasonably used as a measure of conversion efficiency, as in Ref. [45]. In a multibeam interaction system, however, this quantity should be interpreted with caution. For example, in a double-$\Lambda$ four-wave-mixing system, Eq. (B1) corresponds to a conversion efficiency only when the two strong control fields are equal, so that absorption and gain are perfectly balanced, as discussed in Refs. [50-52]. Otherwise, this quantity may exceed unity, as in the present system, whereas a true efficiency cannot exceed unity. Because this quantity compares the generated signal-field intensity with the incident probe-field intensity, it measures the amplification of the generated signal field relative to the probe field. In the system considered here, the strong control field $\Omega_c$ pumps population from level $|2\rangle$ to the upper level $|3\rangle$, thereby providing additional energy for signal-field generation. Therefore, $G_s > 1$ is physically allowed and indicates amplification of the signal field rather than a conversion efficiency exceeding 100%. Accordingly, this quantity is more appropriately interpreted as the gain of the generated signal field.

Following Ref. [71], we define the conversion efficiency more rigorously as

$$\eta_s(z) = \frac{|\Omega_s(z)|^2}{|\Omega_p(0)|^2 + |\Omega_c(0)|^2}, \tag{B2}$$

Unlike $G_s$, the quantity $\eta_s(z)$ normalizes the generated signal by the combined input intensity of the probe and control fields. It describes the ratio of the generated signal-field intensity to the total input-field intensity; because this ratio does not exceed unity, $\eta_s(z)$ can be used as a rigorous measure of efficiency.

The relation between the signal gain and the conversion efficiency follows directly from Eqs. (B1) and (B2):

$$\eta_s(z) = \frac{G_s(z)}{1 + |\Omega_c(0)/\Omega_p(0)|^2}. \tag{B3}$$

Eq. (B3) shows that the signal gain and the conversion efficiency characterize different aspects of the nonlinear process. A larger $G_s$ indicates that the generated signal field is stronger relative to the incident probe field. By contrast, a larger $\eta_s(z)$ requires the generated signal-field intensity to constitute a large fraction of the total intensity of the probe and control fields.

In the parameter regime considered in this work, $|\Omega_c(0)| >> |\Omega_p(0)|$, so Eq. (B3) can be approximated as

$$\eta_s(z) \approx G_s(z) \left|\frac{\Omega_p(0)}{\Omega_c(0)}\right|^2, \tag{B4}$$

Therefore, a strong control field can produce a large signal-field gain even when the normalized efficiency remains relatively small. This distinction is particularly important when comparing the EIT and ATS mechanisms, because operation in the ATS regime generally involves a stronger control field.

## Appendix C: Gain advantage of the ATS scheme over the EIT scheme in a broad parameter range

As shown in Figs. 6(b) and 6(d), when $|\Omega_c|$ is not sufficiently strong, the signal field gain at the exit of the medium in the ATS region can be smaller than that in the EIT region. This behavior arises from the competition among coherent frequency conversion, absorption, and the finite propagation length. Nevertheless, this local behavior does not contradict the gain advantage of the ATS regime under optimized operating conditions. To quantify this distinction, we compare the maximum signal gain attainable in the ATS region with that attainable in the EIT region. To this end, we define the output signal-field gain as

$$G_s(\Omega_c;\lambda)=\left|\frac{\Omega_s(L)}{\Omega_p(0)}\right|^2, \tag{C1}$$

where $\lambda$ denotes the set of parameters other than the control field, namely $\lambda=\{\alpha,\Gamma_{21},\Gamma_{31},\Gamma_{32},\Delta_p\}$. The critical control-field strength separating the EIT and ATS regimes is given by

$$\Omega_{th}=\frac{1}{2}\left|\Gamma_3-\Gamma_{21}\right|. \tag{C2}$$

Because the control-field strength considered in this work is limited to $|\Omega_c|=2\Gamma_3$, we define the maximum gain attainable in the EIT regime as

$$G_{EIT}^{\max}(\lambda)=\max_{0<|\Omega_c|<\Omega_{th}} G_s(\Omega_c;\lambda), \tag{C3}$$

and the maximum gain attainable in the ATS regime as

$$G_{ATS}^{\max}(\lambda)=\max_{\Omega_{th}<|\Omega_c|<2\Gamma_3} G_s(\Omega_c;\lambda). \tag{C4}$$

Accordingly, we define the ratio between the maximum gains attainable in the ATS and EIT regimes as

$$A_{ATS/EIT}=\frac{G_{ATS}^{\max}}{G_{EIT}^{\max}}, \tag{C5}$$

when $A_{ATS/EIT}>1$, the optimized gain attainable in the ATS regime is strictly larger than that attainable in the EIT regime under the same medium parameters and over the same control field scanning range. To distinguish the maximum performance achievable through the two mechanisms, Fig. C1 shows a parameter map of the maximum gain ratio between the ATS and EIT regimes. The color scale represents $\log_{10} A_{ATS/EIT}$. Consequently, the red regions correspond to $G_{ATS}^{\max}>G_{EIT}^{\max}$, whereas the blue regions correspond to $G_{ATS}^{\max}<G_{EIT}^{\max}$. The solid black contour marks the boundary $G_{ATS}^{\max}=G_{EIT}^{\max}$, and the yellow star indicates the parameters used in Fig. 4(b). Fig. C1(a) shows the dependence of the optimized gain ratio on the probe detuning $\Delta_p$ and optical depth $\alpha$. At small optical depths and near resonance, the maximum EIT gain can be comparable to or larger than the maximum ATS gain, as indicated by the blue region enclosed by the black contour. In this short-interaction-length region, the stronger ATS control field does not have sufficient propagation distance to build up its maximum conversion advantage, whereas the interference-assisted EIT process can still produce appreciable signal generation. At resonance, the crossover occurs at approximately $\alpha\approx 6.95$. Above this optical depth, the optimized ATS gain exceeds the optimized EIT gain. As the optical depth increases, most of the investigated detuning range becomes red, demonstrating that the ATS advantage becomes increasingly robust with propagation length. For the optical depth used in the paper, $\alpha=30$, the optimized ATS gain remains higher than the optimized EIT gain throughout the investigated interval $-2\Gamma_3\le\Delta_p\le 2\Gamma_3$.The increase in the ATS-to-EIT ratio away from resonance indicates that the maximum EIT gain is more sensitive to the probe detuning, because detuning weakens the destructive interference responsible for the EIT response. By contrast, the strong-control dressed-state coupling in the ATS regime retains a larger relative signal-generation capability. The weak nonmonotonic structures near resonance at large optical depth arise from the propagation-dependent coherent exchange between the probe and signal fields. Because the gain is evaluated at the fixed output position $z = L$, this coherent exchange produces alternating enhancement and suppression as $\alpha$ varies. Fig. C1(b) shows the combined effects of the lower-state decay rate $\Gamma_{21}$ and optical depth $\alpha$. At small $\Gamma_{21}$ and small $\alpha$, the long-lived coherence associated with the EIT process allows EIT to compete effectively with ATS, producing the blue region in the lower-left part of the map. Increasing either $\Gamma_{21}$ or $\alpha$ moves the system across the black contour into the red region. A larger $\Gamma_{21}$ suppresses the interference-assisted EIT gain more strongly, while the ATS process remains supported by the dressed-state splitting produced by the strong control field. Meanwhile, increasing $\alpha$ provides a longer effective interaction length over which the ATS-mediated signal field can accumulate. The downward shift of the crossover optical depth with increasing $\Gamma_{21}$ therefore shows that the relative ATS advantage becomes more pronounced when the EIT coherence is more strongly affected by decay. For the parameter set considered in this work, namely $\Gamma_{21} = 0.05\Gamma_3$ and $\alpha = 30$, the system lies well within the ATS-dominant region. At this point, $G_{ATS}^{\max} = 0.948$ and $G_{ATS}^{\max} = 7.986$, which gives $A_{ATS/EIT} = 8.42$. Thus, the maximum signal gain attainable in the ATS regime is more than eight times the maximum attainable in the EIT regime for the parameters used in Fig. 4(b). The parameter maps in Fig. C1 further show that this optimized ATS advantage persists over a broad range of optical depths, detunings, and decay-rate combinations.

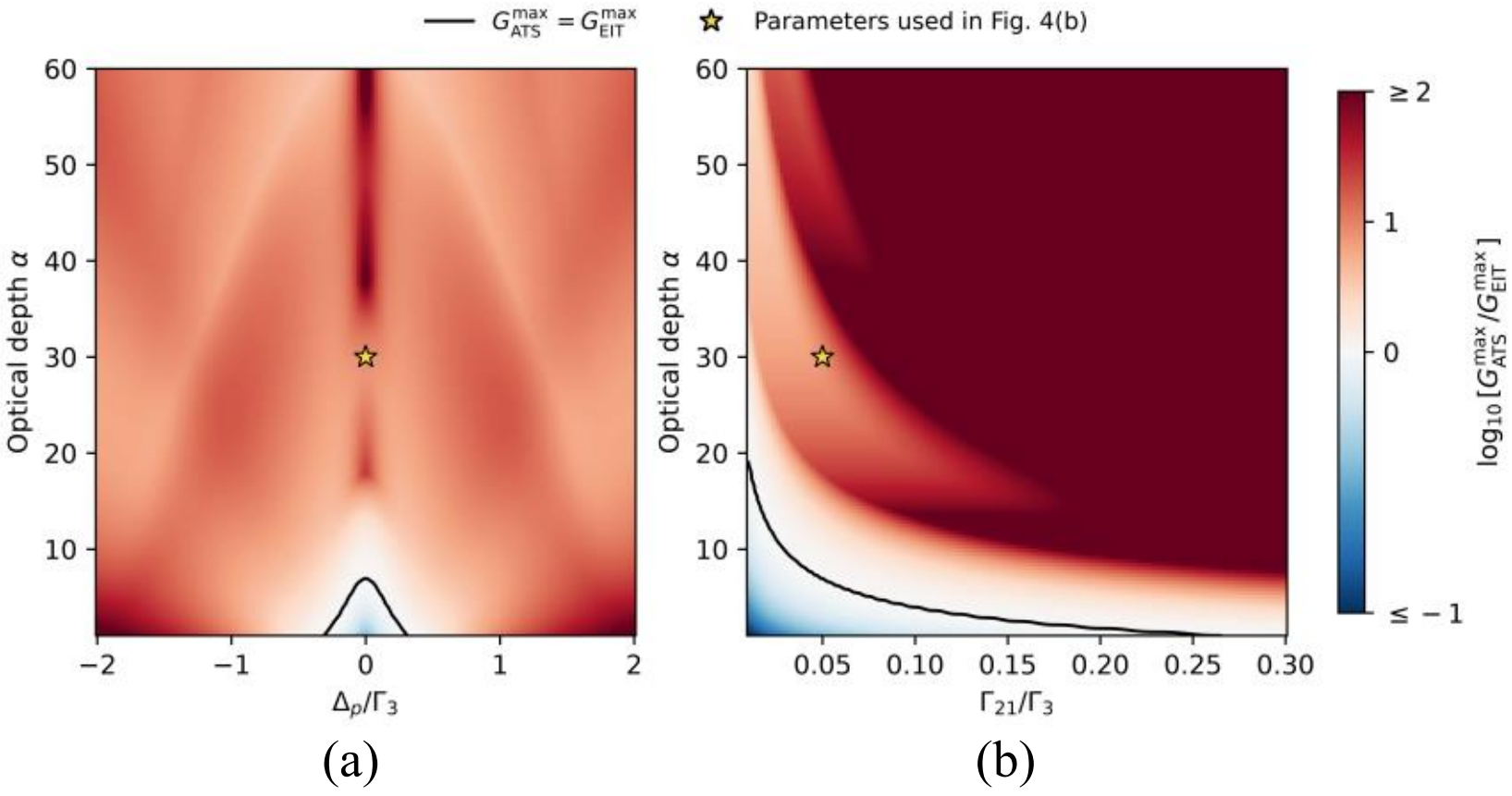

(a) (b)

**Fig. C1.** Comparison between the maximum output signal-field gains attainable in the ATS and EIT regimes. Red regions indicate that the maximum ATS gain exceeds the maximum EIT gain, whereas blue regions indicate the opposite. The black contour represents $G_{ATS}^{\max} = G_{EIT}^{\max}$ .The yellow star marks the parameters used in Fig. 4(b). (a) Dependence on the probe detuning and optical depth, with $\Gamma_{21} = 0.05\Gamma_3$ . (b) Dependence on $\Gamma_{21}$ and optical depth, with $\Delta_p = 0$ . For all panels, $\Omega_{p0} = 0.01\Gamma_3$, $\Delta_c = 0$, and $z = L$.

## Disclosures

The authors declare no conflicts of interest.

## Acknowledgements

We thank Professors Majed S. Fataftah and Danna E. Freedman for insightful and productive discussions. This work was supported by the National Natural Science Foundation of China (Grant Nos. 12474353 and 12474354).